\documentstyle[eqsecnum,aps,epsf,pre]{revtex}
\begin{document}
%_____________________________________________________________________________
\title{Crisis-induced intermittency in truncated mean field dynamos}
%_____________________________________________________________________________
\author{Eurico Covas\thanks{E-mail:E.O.Covas@qmw.ac.uk} and
Reza Tavakol\thanks{E-mail:reza@maths.qmw.ac.uk}}
%_____________________________________________________________________________
\address{Astronomy Unit         \\
School of Mathematical Sciences \\
Queen Mary \& Westfield College \\
Mile End Road                   \\
London E1 4NS, UK}
%_____________________________________________________________________________
\date{\today}
%_____________________________________________________________________________
\maketitle
%_____________________________________________________________________________
\begin{abstract}
%_____________________________________________________________________________
We investigate the detailed dynamics of a truncated $\alpha\omega$
dynamo model with a dynamic $\alpha$ effect. We find the presence of
multiple attractors, including two chaotic attractors with a fractal
basin boundary which merge to form a single attractor as the control
parameter is increased. By considering phase portraits and the scaling
of averaged times of transitions between the two attractors,
we demonstrate that this merging
is accompanied by a crisis-induced intermittency. We also find a range
of parameter values over which the system has a fractal parameter
dependence for fixed initial conditions. This is the first time this
type of intermittency has been observed in a dynamo model and it could
be of potential importance in accounting for some forms of intermittency
in the solar and stellar output.
%_____________________________________________________________________________
\end{abstract}
%_____________________________________________________________________________
\section{Introduction}
%_____________________________________________________________________________
Intermittent type behaviour has been observed in a wide range of
experimental and numerical studies of dynamical systems.
Theoretical attempts at understanding such modes of behaviour fall into
two groups: (i) stochastic, involving models in which intermittency is
brought about through the presence of some form of external noise and
(ii) deterministic, where the mechanism of production of intermittency
is purely internal.

Here we concentrate on the latter and in particular on an important
subset of such mechanisms referred to as ``crisis intermittency''
\cite{grebogietal82,grebogietal87}, whereby attractors underlying the
dynamics change suddenly as a system parameter is varied. There are both
experimental and numerical evidence for such modes of behaviour (see
for example \cite{dittoetal89,grebogietal87,karakotsu96,ottbook93} and
references therein). As far as their detailed underlying mechanism and
temporal signature are concerned, crises come in three varieties
\cite{grebogietal87}. Of particular interest for our discussion here is
the type of intermittency (which can occur in systems with symmetry)
referred to as ``attractor merging crisis'', whereby as a system
parameter is varied, two or more chaotic attractors merge to form a
single attractor. 
\\

An important potential domain of relevance of dynamical intermittency is
in understanding the mechanism of production of the so called ``grand
or Maunder type minima'' in solar and stellar activity, during
which the amplitude of the stellar cycle is greatly diminished
\cite{weiss}. Many attempts have recently been made to account for such
a behaviour by employing various classes of models, including truncated
models involving ordinary differential equations (ODE)
(c.f.\ Weiss et al.\ \cite{cwj84}, Zeldovich et al.\ \cite{zeldovich83}, Spiegel
\cite{spiegel})
as well as axisymmetric mean field dynamo models modelled on partial
differential equations (PDE), in both spherical shell
\cite{offdynamo,tobias,tt} and torus \cite{brookemoss} topologies. In
order to transcend phenomenological explanations and establish the
underlying mechanism for such behaviour\footnote{or behaviours, since
after all more than one intermittency mechanism may 
occur even in a single model but at different system parameters.}, 
it is of vital importance to
be able to distinguish between the various intermittency mechanisms and
this in turn is greatly assisted by 
determining the forms of intermittency that
can occur for stellar dynamo models.

Here we consider a truncation of an axisymmetric mean field
dynamo model and demonstrate that it can possess
crisis-induced intermittency. To begin with 
we find that the system possesses multiple
attractors (including two chaotic ones) with fractal basin boundaries,
over a wide range of control parameters. We also find parameter intervals
over which the system has fractal parameter
dependence for fixed initial conditions. Such fractal structures
can give rise to a form of fragility (final state sensitivity),
whereby small changes in the initial state or the control parameters of
the system can result in a different final outcome. We find
parameter regions where as the control parameter is varied, the chaotic
attractors merge into one attractor thus resulting in crisis-induced
intermittency. We verify this by investigating the phase space of the
system and calculating the scaling exponent put forward by Grebogi et
al. \cite{grebogietal87}. As far as we are aware, this is the first
example of such behaviour in a dynamo model as well as in a
6--dimensional flow.
\\

The structure of the paper is as follows. In section 2 we briefly
introduce the model. Section 3 summarizes our results demonstrating the
presence of crisis in this model and finally section 4 contains our
conclusions.

%_____________________________________________________________________________
\section{The model}
%_____________________________________________________________________________
The dynamo model considered here is the so called
$\alpha \omega$ mean field dynamo model with a 
dynamic $\alpha$--effect given by
Schmalz \& Stix \cite{schmalzetal91} (see also Covas
et al.\ \cite{covasetal96} for details). We assume 
a spherical axisymmetrical configuration with one spatial
dimension $x$ (measured in terms of the stellar radius $R$) for which
the magnetic field takes the form
%_____________________________________________________________________________
\begin{equation}\label{Bspherical}
\vec{B}=\left(0,B_{\phi},\frac{1}{R}\frac{\partial A_{\phi}}{\partial x}\right),
\end{equation}
%_____________________________________________________________________________
where $A_\phi$ is the $\phi$--component (latitudinal) of the magnetic
vector potential and $B_\phi$ is the $\phi$--component of $\vec{B}$.
The model is made up of two ingredients:
%_____________________________________________________________________________
\begin{description}
%_____________________________________________________________________________
\item[(I)] the mean field induction equation
%_____________________________________________________________________________
\begin{equation}\label{induction}
\frac{\partial\vec{B}}{\partial t}=\nabla\,\times\,(\vec{v}\,\times\,\vec{B}+
\alpha\vec{B}-\eta_t\nabla\,\times\,\vec{B}),
\end{equation}
%_____________________________________________________________________________
where $\vec{B}$ is the mean magnetic field, $\vec{v}$ is the mean
velocity, $\eta_t$ is the turbulent magnetic diffusitivity
and $\alpha$ represents the $\alpha$--effect.
%_____________________________________________________________________________
\\

\item[(II)] The 
$\alpha$--effect which 
arises from the correlation of small scale turbulent
velocity and magnetic fields\cite{krause80} and is
important in maintaining the dynamo action by relating the mean
electrical current arising in helical turbulence to the mean magnetic
field. Here $\alpha$ 
is assumed to be dynamic and expressible in the form
$\alpha=\alpha_0\cos x-\alpha_M(t)$, where $\alpha_0$ is a constant and
$\alpha_M$ is its dynamic part satisfying the equation
%_____________________________________________________________________________
\begin{equation}\label{dynamicalpha}
\frac{\partial \alpha_M}{\partial t}= \nu_t
\frac{\partial^2 \alpha_M}{\partial x^2} + Q\,\vec{J}\cdot\vec{B},
\end{equation}
%_____________________________________________________________________________
where $Q$ is a physical constant, $\vec{J}$ is the electrical current
and $\nu_t$ is the turbulent diffusivity.
\end{description}
%_____________________________________________________________________________
These assumptions allow Eq. (\ref{induction}) to be split into the
following two equations:
%_____________________________________________________________________________
\begin{eqnarray}
\label{p1}
\frac{\partial A_{\phi}}{\partial t}&=&\frac{\eta_t}{R^2}
\frac{\partial^2A_{\phi}}{\partial x^2}+\alpha B_{\phi},\\
\label{p2}
\frac{\partial B_{\phi}}{\partial t}&=&
\frac{\eta_t}{R^2}\frac{\partial^2 B_{\phi}}
{\partial x^2}+\frac{\omega_0}{R}\frac{\partial A_{\phi}}{\partial x}.
\end{eqnarray}
%_____________________________________________________________________________
Expressing these equations in a
non-dimensional form, relabelling the new variables thus
%_____________________________________________________________________________
\begin{equation}
(A_\phi,~B_\phi,~ \alpha_M) \Longrightarrow (A,~B,~C),
\end{equation}
%_____________________________________________________________________________
and using a spectral expansion of the form
%_____________________________________________________________________________
\begin{eqnarray}
A=\sum_{n=1}^{N}A_n(t)\sin nx,\\
B=\sum_{n=1}^{N}B_n(t)\sin nx,\\
C=\sum_{n=1}^{N}C_n(t)\sin nx,
\end{eqnarray}
%_____________________________________________________________________________
where $N$ determines the truncation order, reduces the equations
(\ref{dynamicalpha}), (\ref{p1}) and (\ref{p2}) into a set of ODE, the
dimension of which depends on the truncation order $N$. In Covas et
al.\ \cite{covasetal96}, the models were taken to be antisymmetric with
respect to the equator and it was found that the minimum truncation
order $N$ for which a similar asymptotic behaviour existed was
$N=4$. Here in view of computational costs, we take this value of $N$
for which the set of truncated equations becomes:
%\widetext
%_____________________________________________________________________________
\begin{eqnarray}
%_____________________________________________________________________________
\label{tr1}
\frac{\partial A_1}{\partial t}&=&
-A_{{1}}+{\frac {D B_{{2}}}{2}}-{\frac {32\,B_{{2}}C_{{2}}}{15\,\pi 
}}+{\frac {64\,B_{{2}}C_{{4}}}{105\,\pi }}+{\frac {64\,B_{{4}}C_{{2
}}}{105\,\pi }}-{\frac {128\,B_{{4}}C_{{4}}}{63\,\pi }}\\
%_____________________________________________________________________________
\label{tr2}
\frac{\partial B_2}{\partial t}&=&
-4\,B_{{2}}+{\frac {8\,A_{{1}}}{3\,\pi }}-{\frac {24\,A_{{3}}}{5\,
\pi }}\\
%_____________________________________________________________________________
\label{tr3}
\frac{\partial C_2}{\partial t}&=&
-4\,\nu\,C_{{2}}+{\frac {16\,A_{{1}}B_{{2}}}{5\,\pi }}-{\frac {32\,
A_{{1}}B_{{4}}}{7\,\pi }}+{\frac {144\,A_{{3}}B_{{2}}}{7\,\pi }}+{
\frac {416\,A_{{3}}B_{{4}}}{15\,\pi }}\\
%_____________________________________________________________________________
\label{tr4}
\frac{\partial A_3}{\partial t}&=&
-9\,A_{{3}}+{\frac {D B_{{2}}}{2}}+{\frac {D B_{{4}}}{2}}-{\frac {32
\,B_{{2}}C_{{2}}}{21\,\pi }}-{\frac {64\,B_{{2}}C_{{4}}}{45\,\pi }}
-{\frac {64\,B_{{4}}C_{{2}}}{45\,\pi }}-{\frac {128\,B_{{4}}C_{{4}}
}{165\,\pi }}\\
%_____________________________________________________________________________
\label{tr5}
\frac{\partial B_4}{\partial t}&=&
-16\,B_{{4}}+{\frac {16\,A_{{1}}}{15\,\pi }}+{\frac {48\,A_{{3}}}{7
\,\pi }}\\
%_____________________________________________________________________________
\label{tr6}
\frac{\partial C_4}{\partial t}&=&
-16\,\nu\,C_{{4}}+{\frac {96\,A_{{1}}B_{{2}}}{35\,\pi }}+{\frac {64
\,A_{{1}}B_{{4}}}{21\,\pi }}+{\frac {32\,A_{{3}}B_{{2}}}{3\,\pi }}+
{\frac {576\,A_{{3}}B_{{4}}}{55\,\pi }},
%_____________________________________________________________________________
\end{eqnarray}
%\narrowtext
where $D$ is the control parameter, the so called dynamo number, and
$\nu=\frac{\nu_t}{\eta_t}$ which for compatibility with
\cite{covasetal96,schmalzetal91} we take to be $\nu=0.5$.

Clearly the details of the resulting dynamics will depend on the
truncation order chosen. For example, the $N=2$ case is expressible as the
3--dimensional Lorenz system and the higher truncations can have
different quantitative types of behaviour. The important point, as far as
our discussion here is concerned, is that the multi-attractor 
regime discussed here
seems to be present as the order of truncation is
increased. In this way such a behaviour might be of
potential relevance in understanding some of the intermittent behaviour
in the output of the Sun and other stars.
%_____________________________________________________________________________
\section{Crisis-induced intermittency}
%_____________________________________________________________________________
A coarse study of the system (\ref{tr1}) -- (\ref{tr6}) and higher
truncations was reported in \cite{covasetal96} from a different point of
view. Here we demonstrate the occurrence of crisis-induced intermittency in
this system by considering the detailed nature of its attractors, their
basins and especially their metamorphoses (merging), while treating $D$
as the control parameter. 

To begin with we recall that symmetries are usually 
associated with this type of attractor merging. The six dimensional
dynamical system considered here possesses the symmetries:
%_____________________________________________________________________________
\begin{equation}
A_n \to -A_n, \quad B_n \to -B_n, \quad C_n \to C_n.
 \end{equation}
%_____________________________________________________________________________

Now assuming the existence of a crisis for this system at $D=D_c$, then for
crisis-induced intermittency to exist one requires that for $D<D_c$
there exist two (or more) chaotic attractors and that as $D$ is
increased, the attractors enlarge and at $D=D_c$ they simultaneously
touch the boundary separating their basins. In that case, for $D$
slightly greater than $D_c$, a typical orbit will spend long periods of
time in each of the regions where the attractors existed for $D<D_c$ and
intermittently switch between them. An important signature for this
mechanism is the way the average time $\tau$ between these switches
scales with the system parameter $D$. According to Grebogi et
al.\ \cite{grebogietal87}, for a large class of dynamical systems, this
relation takes the form
%_____________________________________________________________________________
\begin{equation}\label{index}
\tau \sim \left|D-D_c \right |^{-\gamma},
\end{equation}
%_____________________________________________________________________________
where the real constant $\gamma$ is the critical exponent characteristic
of the system under consideration.

\begin{figure}
\centerline{\def\epsfsize#1#2{0.4#1}\epsffile{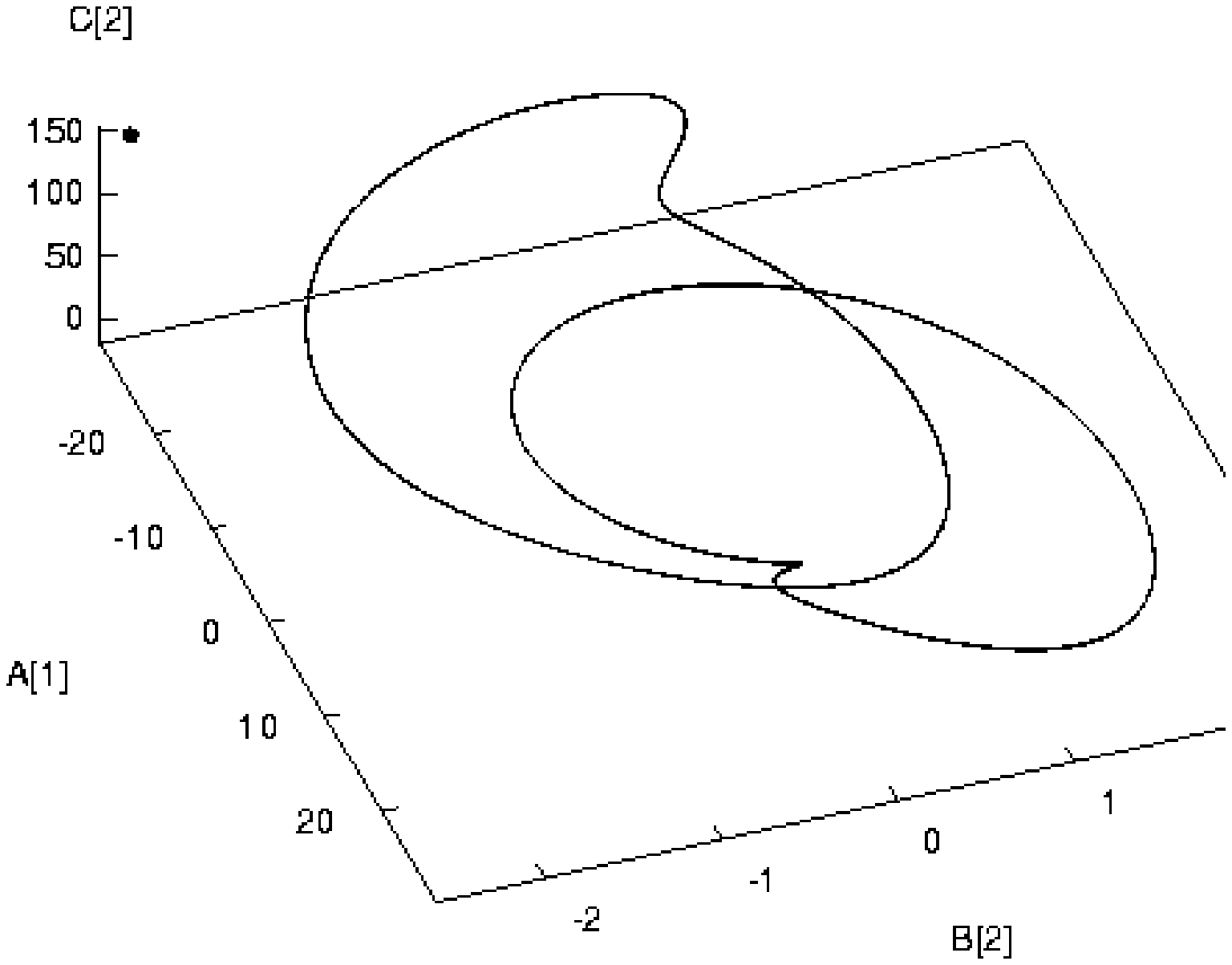}}
\caption{\label{attractors1}
Phase portraits of the two fixed points and the two stable cycles}
\end{figure}

\begin{figure}
\centerline{\def\epsfsize#1#2{0.4#1}\epsffile{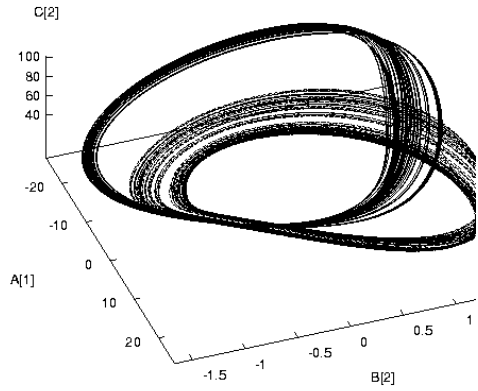}}
\caption{\label{attractors2}
Phase portraits of the two coexistent chaotic attractors}
\end{figure}

To show that crisis-induced intermittency 
occurs for the system (\ref{tr1}) -- (\ref{tr6}), we
begin by noting that our numerical results indicate that, for a wide range of
parameter values, the system possesses multiple attractors consisting of
fixed points, periodic orbits and chaotic attractors. Starting around
$D=195$, two cycles coexist and both bifurcate in a doubling bifurcation
sequence into two chaotic attractors that coexist after $D>203$. At $D
\approx 200.4$ two other periodic orbits appear which persist for the
parameter values considered here. Figures \ref{attractors1} and
\ref{attractors2} show these attractors for
$D=204$, where all 6 coexist and their positions in the 6--dimensional
phase are well separated (note that the apparent overlaps in Figs
\ref{attractors1} and \ref{attractors2} are due to projections).

\begin{figure}
\centerline{\def\epsfsize#1#2{0.28#1}\epsffile{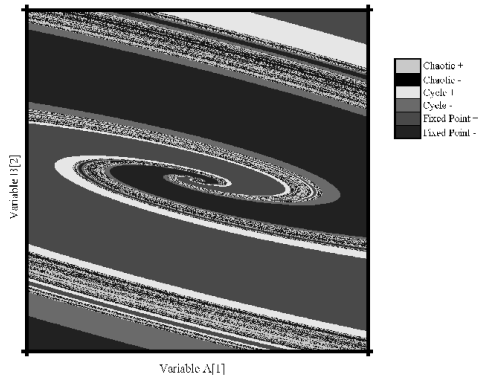}}
\caption{\label{basins}
A $800 \times 800$ grid showing a 
2--D cut of the basins of attraction with $D=204$ and
$C_2=A_3=B_4=C_4=0$. Variables $A_1$ and $B_2$ were centred at $(0,0)$
and the size of the picture is 2 by 1. In the legend, $+$ and $-$
indicate the sign of the time average of $A_1$}
\end{figure}

\begin{figure}
\centerline{\def\epsfsize#1#2{0.28#1}\epsffile{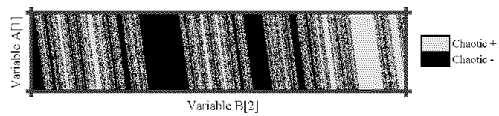}}
\caption{\label{magnify}
A $800 \times 160$ grid showing the amplification of the previous
picture (close to the lower left corner) with $A_1=-0.804$ and
$B_2=-0.700$ and with size 0.01 by 0.002}
\end{figure}

We also found the corresponding basins of attraction for each attractor
which indicate fractal boundaries. This can be seen in Figure
\ref{basins} which shows a two dimensional cut $(C_2=A_3=B_4=C_4=0)$ of
the basin boundary for this system at the parameter value $D=204$ and
Figure \ref{magnify} which shows the magnification of a region of Figure
\ref{basins} where both chaotic attractors possess fractal basins \cite{html}.
We also calculated the box counting dimension
of the boundary between attractors on a horizontal 1--D cut of Figure
\ref{magnify}, which turned out to be non integer, further
substantiating the fractal nature of the boundaries.

\begin{figure}
\centerline{\def\epsfsize#1#2{0.4#1}\epsffile{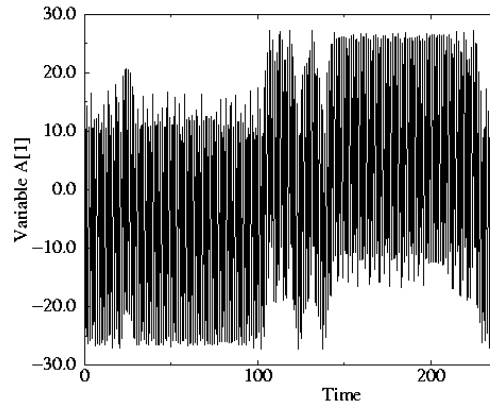}}
\caption{\label{series} Chaotic time series for the merged attractors
for $D=205>D_c$}
\end{figure}

\begin{figure}
\centerline{\def\epsfsize#1#2{0.4#1}\epsffile{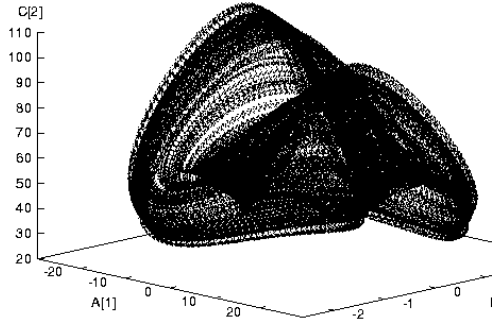}}
\caption{\label{merged} The projection of the resulting merged chaotic
attractor in the space $A_1, B_2, C_2$ for $D=207$}
\end{figure}

Now as $D$ is increased, the two chaotic attractors merge and give rise
to a single connected attractor. Figure \ref{series} shows
the time series for the variable
$A_1$ after the merging and Figure \ref{merged} shows the
projection of the merged attractors on the variables $A_1$, $B_2$ and
$C_2$.
Prior to $D_c\approx 204.2796$, there is no switch
between the two attractors and the time series does not show the 
bimodal behaviour seen in Figure \ref{series}.

\begin{figure}
\centerline{\def\epsfsize#1#2{0.4#1}\epsffile{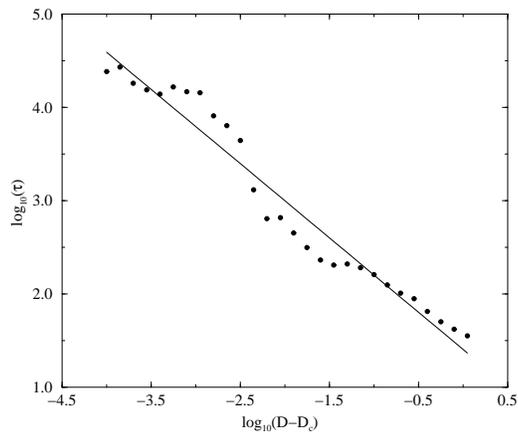}}
\caption{\label{scaling}
Scaling of $\tau$ as a function of the distance to the critical dynamo
number $D_c$ together with the fitted line}
\end{figure}

These results show a clear indication for the occurrence of
crisis-induced intermittency in this model. To substantiate this
further, we checked that for this system the scaling relation
(\ref{index}) is satisfied in the neighbourhood of $D_c\approx
204.2796$. Figure \ref{scaling} shows the plot of $\log_{10}|\tau|$
versus $\log_{10}\left|D-D_c\right|$. To produce the plot, 28 points were taken at
regular spacings with the initial conditions chosen in the chaotic basin of the
merged attractor after $D\approx 204.2796$ and 200 million iterations were taken
for each point.
The transitions between the ghosts of the previous attractors were detected
using the averages of the variable $A_1$ over a pseudo-period of
approximately $\Delta t\approx 1.5$ non-dimensional time units. As can
be seen the points are well approximated by a straight line, which was
obtained using a least squares fit which giving $\gamma
\approx 0.79 \pm 0.03$.

The $\gamma$ coefficient can be calculated also from theoretical grounds,
as shown in Grebogi et al.\ \cite{grebogietal87}. The method involves
calculating the stable and unstable manifolds of the unstable orbit (thereafter $B$) 
mediating the crisis. By examining the trajectories around the transitions
between the ghosts of the previous attractors at $D=204.35>D_c$, we found 
the point where
the orbit went inside the portion of the unstable manifold of the
$B$ that has poked over to the other side
of the stable manifold of $B$. The orbit then follows closely the orientation of 
the stable and unstable manifolds. We then calculated a  estimate of
the direction of the unstable and stable manifolds. Since this
was very sensitive, the value of $\gamma$ had a large error bar, 
that is, the calculated value could be anywhere on the range $[0.4,1.2]$,
depending on minor changes in the choice of the vectors that determine
the unstable and stable manifolds. Because the system was high dimensional,
all the projections in two dimensional planes we used were not very
useful to determine with good precision the directions of the two
manifolds. Therefore we were unable to calculate
the critical exponent with sufficient precision to compare with
the one calculated from the time between flips of the orbit.  

\begin{figure}
\centerline{\def\epsfsize#1#2{0.5#1}\epsffile{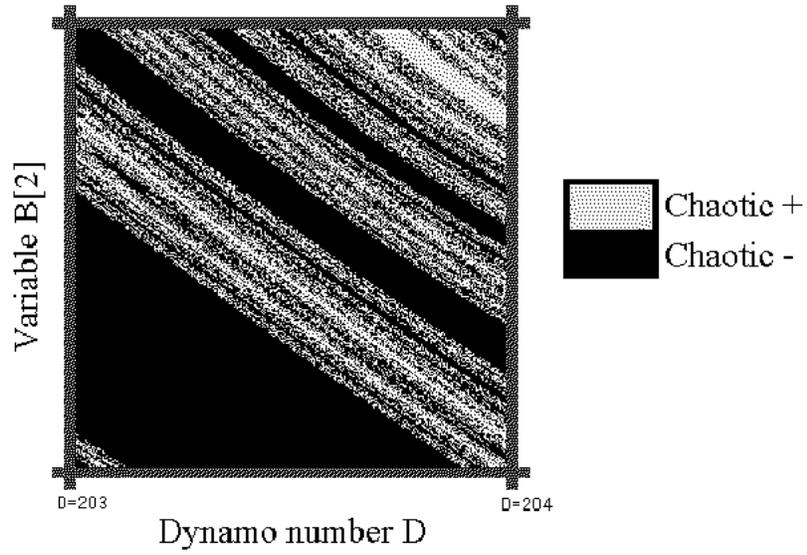}}
\caption{\label{parameter}
Depiction of the final state (attractor) of the system as a function of
changes in the parameter $D$ and the initial condition $B_2$. This
Figure represents a horizontal slice of Fig. \ref{magnify} for many runs
with different dynamo numbers. A resolution of 300 by 300 pixels was
used and all initial conditions were taken to be 
zero except for $A_1=-0.80$ and $B_2$
centred at -0.70}
\end{figure}

Finally we looked at the parameter dependence of the system
for fixed initial conditions.
We found that there are intervals of $D$
for which this is fractal. This can be seen from 
Figure \ref{parameter} which depicts the
final state (attractor) of the system (\ref{tr1}) -- (\ref{tr6})
as a function of changes in the parameter $D$
and the initial condition $B_2$.

%_____________________________________________________________________________ 
\section{Conclusions}
%_____________________________________________________________________________
We have found the presence of multiple attractors with fractal basin
boundaries as well as crisis-induced intermittency in a truncated
axisymmetric $\alpha \omega$ dynamo model which is antisymmetric with
respect to the equator. We have seen that this type of intermittency is
due to the collision of the two chaotic attractors
and have confirmed this by calculating the scaling coefficient suggested
by Grebogi et al.\ \cite{grebogietal87}.

The presence of crisis-induced intermittency, coupled with the facts that 
this type of multiple attractors
seem to persist in higher order truncations
and the presence of symmetry in dynamo models,
may indicate the relevance of this type of
intermittency in more realistic dynamo settings.

We have also found that this system possesses 
fractal parameter dependence for fixed initial conditions.
The presence of such fractal structures 
results in a form of fragility (final state sensitivity), 
whereby small changes in the initial
conditions or the control parameter of the system 
can result in qualitative changes in its final dynamics. This
type of sensitivity could be of significance in astrophysics in that, for
example, it could potentially lead to stars of same spectral type,
rotational period, age and compositions showing different modes of
dynamical behaviour \cite{tt95}.

Finally as far as we are aware, this is the first instance of 
such behaviour in a dynamo model as well as in a $6D$ flow. 

\acknowledgments 

We would like to thank John Brooke and Andrew Tworkowski
for helpful discussions. We also thank an anonymous referee
for his useful comments and criticisms. 
EC is supported by grant BD / 5708 / 95 -- Program PRAXIS
XXI, from JNICT -- Portugal. RT benefited from PPARC UK Grant No.
H09454. This research also benefited from the EC Human Capital and
Mobility (Networks) grant ``Late type stars: activity, magnetism,
turbulence'' No. ERBCHRXCT940483.

%_____________________________________________________________________________

\pagebreak
%_____________________________________________________________________________
%_____________________________________________________________________________
%_____________________________________________________________________________
%_____________________________________________________________________________
%_____________________________________________________________________________
%_____________________________________________________________________________
%_____________________________________________________________________________
%_____________________________________________________________________________
%_____________________________________________________________________________
\end{document}